\documentclass{jpsj-suppl}
\usepackage{txfonts} 

\title{Combined Partial Wave Analysis for the description of
exclusive p+p $\rightarrow$ p+K$^+$+$\Lambda$  production}

\author{Robert Muenzer$^{1}$}

\inst{$^{1}$Excellence Cluster Universe - Technical University Munich, Boltzmannstrasse 2, 85748 Garching, Germany}

\email{rmuenzer@ph.tum.de}

\recdate{}

\abst{The production of \pkl in elementary p+p collision was investigated using the Bonn Gatchina Partial Wave Analysis framework. This approach allows the determination of possible participating production wave depending on the quantum numbers of the system. For the analysis seven data samples, measured at different detectors and beam energies, were used. For the extraction of the $\Lambda$-p scattering length a cross check with established methods is required. Furthermore the total crosssection of the production process is needed to be determined to extra cross section for the separate waves. Both methods are described in this work.}

\kword{Partial Wave Analysis,Strangeness Production,Lambda-Proton Interaction, \nstar Resonances, $\Sigma$-N Cusp}

\newcommand{\lamo}{$\Lambda$}

\newcommand{\Kpo}{$K^{+}$}

\newcommand{\nstaro}{$N^{*+}$}
\newcommand{\nstar}{\nstaro\ }
\newcommand{\pKLo}{p\Kpo\lamo}

\newcommand{\pklo}{\pKLo}

\newcommand{\pkL}{\pKLo\ }
\newcommand{\pkl}{\pKLo\ }
\newcommand{\distoo}{DISTO}

\newcommand{\disto}{\distoo\ }

\newcommand{\DISTO}{\disto\ }
\newcommand{\cosyo}{COSY-TOF}

\newcommand{\cosy}{\cosyo\ }

\newcommand{\COSY}{\cosy}
\newcommand{\cosytof}{\cosy}

\newcommand{\equ}[1]{Equation~\ref{#1}}

\newcommand{\Equ}[1]{Equation~\ref{#1}}

\begin{document}
\setlength{\unitlength}{0.01\columnwidth}
\maketitle
The production of strangeness in elementary proton-proton reaction is an important ingredient
for different fields in hadron physics. On the one hand a suffcient description of the production
mechanism is required from transport model calculation \cite{1}. The main input for the description of
strangeness production are the energy dependent cross section, the branching ratio and non-isotropic
production distributions. Especially at energies in the low GeV range the production of \pkl can
be realized by non-resonant and several resonant channels, for example \nstar , which play a dominant
role \cite{2}.\\
Furthermore a suffcient understanding of the production mechanism, which includes also p-$\Lambda$ final
state interaction, is necessary to look for new exotic production channels like the one of kaonicclusters.
This state, which may have a very broad width \cite{3}, is likely to be visible only by small
deviation from the known channels. For that reason insuffciencies in the description of the production
can lead to wrong interpretation of such kind of deviations \cite{4}.\\
In the last years a campaign was started to analyze the production mechanism using Partial Wave
Analysis methods, to understand the contributions of different possible production processes including
also interference between the wave functions of the different channels.\\
This aim of a combined analysis is to fit exclusive experimental p+p $\rightarrow$ \pkL data from different
experiments using partial wave analysis method, to extract a description of all data sample with on
set of transition amplitudes.\\
The experimental data used for this analysis are listed in Table I. In this table the beam energy the
available used statistics are listed.\\
\begin{table*}[ht]
\begin{minipage}[t]{0.5\textwidth}
    \begin{tabular}{c|cc}
      Experiment  & E$_{\text{Beam}}$ (GeV) &  statistics\\
      \hline
      \DISTO \cite{5} & 2.14 & 121000  \\
      \COSY-TOF \cite{6} & 2.16 & 43662 \\
      \DISTO \cite{5}    & 2.5 & 304000 \\
      \DISTO \cite{5,9}   & 2.85 & 424000  \\
    \end{tabular}
\end{minipage}
\begin{minipage}[t]{0.5\textwidth}
    \begin{tabular}{c|cc}
      Experiment  & E$_{\text{Beam}}$ (GeV) &  statistics\\
	\hline
      FOPI  \cite{7} & 3.1 & 903 \\
      HADES \cite{4}   & 3.50 & 133155  \\
      HADES \cite{4}   & 3.50 & 8155 \
    \end{tabular}
\end{minipage}
  \caption[Available statistics for the reaction p+p$\rightarrow$\pklo.]{List of beam energy and available statistics for the reaction \pkl measured by the \cosytof,
\disto, FOPI and HADES collaborations.}
  \label{tab:samples}
\end{table*}
The extraction of the exclusive events are explained elsewhere \cite{4,5,6,7}.\\
For the Analysis the Bonn-Gatchina Framework is used, which allows to fit uncorrected data. This is
done by weighting the total transition waves with experiment-specific full scale simulation.
In the BG-PWA the treatment of the final system depends on the corresponding channels. For our
analysis the case of non-resonant and resonant production are treated in a different way. A description
of the total parametrization can be found in \cite{4,7,8}.\\
For resonant production the final system consists of \nstar and p, with certain mass M, width $\Gamma$ and
quantum numbers (see Table II) . For the spectral function a relativistic Breit-Wigner formula is used:\\
\begin{equation}
A_{2b}^{\beta}(s) = \frac{1}{(M^{2}-s-\text{i}\Gamma M)},
\label{equ:common:bg-pwa:amplitude:breitwigner}
\end{equation}
\begin{table}[ht]
\begin{minipage}[t]{0.5\textwidth}
 \begin{tabular}{c|c|c|c|c}
      \nstar & $J^{P}$ & Mass & Width & $\Gamma_{K^{+}\Lambda}/\Gamma_{tot}$ (\%) \\
      & & Gev$c^{-2}$ & Gevc$c ^{-2}$ &  \\
      \hline
      1650&$\frac{1}{2}^{-}$ & 1.655 & 0.150 & 3-11\\
      1710&$\frac{1}{2}^{+}$ & 1.710 & 0.100 & 5-25\\
      1720&$\frac{3}{2}^{+}$ & 1.720 & 0.250 & 1-15\\
      1875&$\frac{3}{2}^{-}$ & 1.875 & 0.220 & 4$\pm$2\\
    \end{tabular}
\end{minipage}
\begin{minipage}[t]{0.5\textwidth}
    \begin{tabular}{c|c|c|c|c}
      \nstar & $J^{P}$ & Mass & Width & $\Gamma_{K^{+}\Lambda}/\Gamma_{tot}$ (\%) \\
      & & GeV$c^{-2}$ & GeV$c ^{-2}$ &  \\
      \hline
      1880&$\frac{1}{2}^{+}$ & 1.870 & 0.235 & 2$\pm$1\\
      1895&$\frac{1}{2}^{-}$ & 2.090 & 0.090 & 18$\pm$5\\
      1900&$\frac{3}{2}^{+}$ & 1.900 & 0.250 & 0-10\\
    \end{tabular}
\end{minipage}
  \caption{\nstar resonances included in the Partial Wave Analysis written in the spectroscopic notation with their the mass and the width, taken form   \cite{10}}
  \label{tab:common:bg-pwa:pwainput:datasamples:nstars} 
\end{table}
The production of the $\Sigma$-N cusp is treated as a two particle system of the K+ and the $\Sigma$-N quasi\-particle. According to the different possible quantum numbers (JP=0+ or 1+) \cite{11} separate wave are
added.\\
The parameterization of the cusp is done in two different approaches. A phenomenological approach
was used using a relativistic Breit-Wigner (\equ{equ:common:bg-pwa:amplitude:breitwigner}), which is motivated by a symmetric shape
of the cusp structure in the experimental data. A pole mass of 2.13GeVc$^{-2}$ and width of 20MeVc$^{-2}$
were used. A second approach using Flatte parametrization \cite{11} is also investigated.\\
In case the final state consists of three particles - like for non-resonant production- the $\Lambda$ and the
proton form a two particle p$\Lambda$ -subsystem. The total final state is considered as a two particle system
of a p$\Lambda$ and a K+ particle. The p-$\Lambda$ FSI is parametrized by an effective range parametrization:
\begin{equation}
  A_{2b}^{\beta} = \frac{\sqrt{s_{i}}}{1-\frac{1}{2}r^{\beta}q^{2}a^{\beta}_{p\Lambda} + \text{i}qa^{\beta}_{p\Lambda} q^{2L}/F\left(q,r^{\beta},L\right)},
\label{equ:common:bg-pwa:amplitude:plscattering}
\end{equation}
where q is the relative momentum between the p and $\Lambda$.$\alpha_{p\Lambda}^{\beta}$ is the p$\Lambda$-scattering length and  r$^{\beta}$ is the effective range of the $\Lambda$-p system for the channel $\beta$. F(q,r,L) is the Blatt-Weisskopf factor, which is used for normalization (with F(q,r,L=0)=1) \cite{8}.\\
The values for the scattering length and the effective range can be set as free parameters for the fit,
which allows the extraction of these values from the fitting procedure. A crosscheck can be done using
the scattering length extraction method described in \cite{12}. In this method one divide the p-$\Lambda$ invariant
mass spectrum including FSI by the spectrum produced by pure phase space. This resulting spectrum
( $\frac{1}{PS}\frac{d\sigma}{dm_{pA}}$)  is fitted by the following function in the region $m_{0} = m_{p} +m_{\Lambda}$ and $m_{max}$ = $m_{0}$ + 40MeVc$^{-2}$.
\begin{equation}   
	\frac{1}{PS}\frac{d\sigma}{dm_{p\Lambda}}  = \exp \left[C_{0} +\frac{C_{1}}{m_{p\Lambda}-C_{2}} \right],
        \label{equ:common:bg-pwa:amplitude:scatteringlenght}
\end{equation}
The scattering length can be determined by:
\begin{equation}    
	\alpha\left(C_{1}, C_{2} \right) = -\frac{1}{2} C_{1}\sqrt{\left(\frac{m_{0}^{2}}{m_{p}m_{\Lambda}}\right)\cdot\frac{\left(m_{max}^{2}-m_{0}^{2}\right)}{\left(m_{max}^{2}-C_{2}\right)\cdot\left(m_{0}^{2}-C_{2}\right)^{3}}\hbar c}
        \label{equ:common:bg-pwa:amplitude:scatteringlenght3}
\end{equation}
The crosscheck was performed by choosing certain input parameters ($\alpha_{Singlet}$=-1.9 GeV$c^{-1}$/ $\alpha_{Triplett}$=-
9.14 GeV$c^{-1}$) for the PWA and extracting the resulting mass spectra for the p-$\Lambda$ S-Wave channel.
Since the values are only used for checking the method, they differ from literature values. In Figure 1
the phase-space divided p-$\Lambda$ invariant mass spectra are plotted for the S-wave singlett (left) and the
S-wave triplett (right) wave. the black line indicates the fitted curve of \Equ{equ:common:bg-pwa:amplitude:scatteringlenght}.\\
From the fit the values $\alpha_{Singlett}$=-3.1$\pm$0.2$\pm$1.5GeVc$^{-1}$ and $\alpha_{Triplett}$=-7$\pm$3$\pm$1.5GeVc$^{-1}$ are extracted. The
first error represent the fitting error, while the second originates in the theory \cite{12}.\\
This shows, that the result agrees with the errors of the method.
\begin{figure}
        \begin{center}
	\includegraphics[width=11cm]{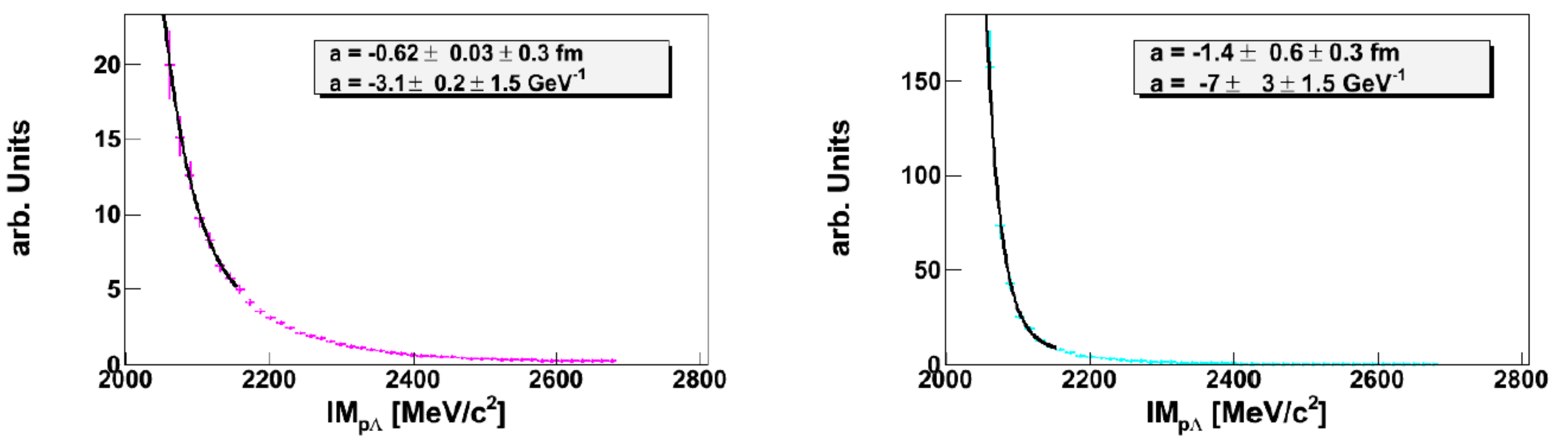}
        \end{center}
	\caption{Invariant mass of $\Lambda$-p distribution, for the non resonant production channel with $\Lambda$-p produced in
S-Wave Singlett (left) and Triplett (right). The spectra are divided by pure phase space distributions.The black
line indicated the fit of the the function given in Equation 3 (see text)}
	\label{fig:combined:hfdca:phasespace}
\end{figure}
After the fitting converges and provides a suffcient description of the experimental data, the relative strength of the different production channels can be extracted. This strength can be translated into
a cross section by multiply these values with the total production cross section of the \pkL at the
corresponding energy.\\
Since the total production cross section was not measured for all experimental data, which are used
in this analysis, the cross section has to be extracted by a Phase-Space function \cite{13}:
\begin{equation}
\sigma_{pKL}[\mu b]=C_{1}\left(1-\frac{s_{0}}{\left(\sqrt{s_{0}}+\epsilon[MeV]\right)^{2}}\right)^{C_{2}}\left(\frac{s_{0}}{\left(\sqrt{s_{0}}+\epsilon[MeV]\right)^{2}}\right)^{C_{3}}
\label{equ:combined:hfdca:phasespace}
\end{equation}
The input value are taken from \cite{2,4,11,14}. The resulting fit function is shown in Figure 2.
The grey band indicated the fitting error. The resulting values from the fit are C1 = 4.0 $\pm$ 0.5 $\cdot$ 10$^{2}$,
C2 = 1,49$\pm$0.04, and C3 = 1.4$\pm$0.4. Using the results from the phase space fit, the total cross section
for the energies of the different data sample can be determined (see Table III). To reduce influence of
systematical error, for all energies the value from the phase space fit are taken.\\
For the calculation of the total production cross section of the\nstar resonances additionally the
branching ratio has to be taken into account, which are taken listed in Table II.

\begin{figure}
        \begin{center}
	\includegraphics[width=9cm]{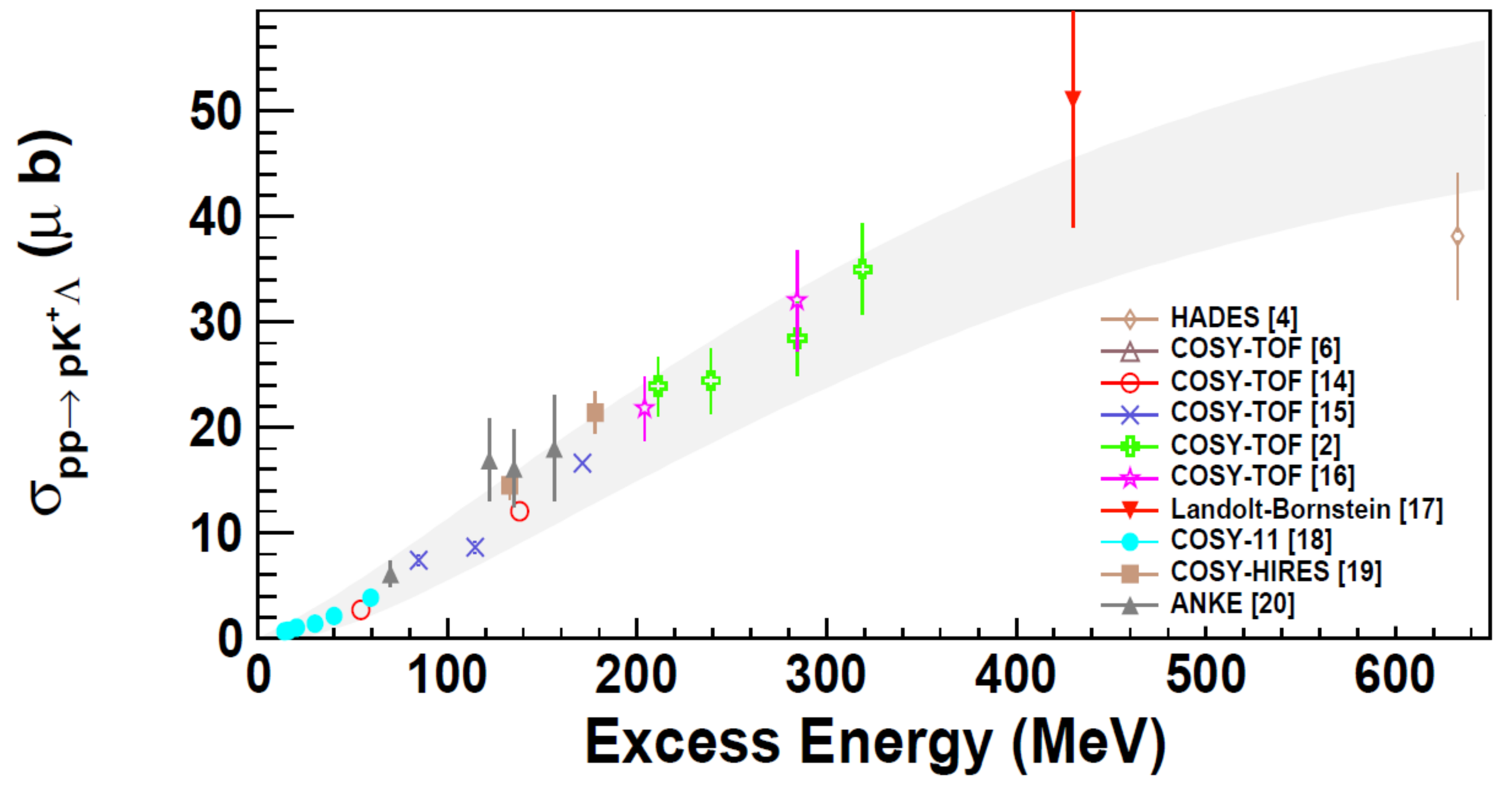}
        \end{center}
	\caption{Total cross section versus excess energy taken from literature values \cite{ 2,4,11,14} ). The grey band lines indicates a phase space it (\equ{equ:combined:hfdca:phasespace})}
	\label{fig:combined:hfdca:phasespace}
\end{figure}
\begin{table*}[ht]
\begin{minipage}[t]{0.5\textwidth}
\begin{center}
    \begin{tabular}{c|cc}
      experiment  & E$_{\text{Beam}}$ &$\sigma_{tot}$\\
      \hline
      \DISTO & 2.14 &$19.0\pm3.3$ \\
      \COSY  & 2.16& $19.7\pm3.5$\\
      \DISTO& 2.5 &$30.5\pm5.7$\\
    \end{tabular}
\end{center}
\end{minipage}
\begin{minipage}[t]{0.5\textwidth}
    \begin{tabular}{c|cc}
      experiment  & E$_{\text{Beam}}$ &$\sigma_{tot}$\\
      \hline
      \DISTO  & 2.85 &$38.7\pm7.9$ \\
      FOPI & 3.1 &$43.1\pm9.3$\\
      HADES & 3.50 &$48.0\pm11.5$ \\
    \end{tabular}
\end{minipage}
  \caption{Total Production cross section of \pkl extracted from Phase-Space fit (see text)}
  \label{tab:samples}
\end{table*}
\paragraph*{Summary}
Using the BG-PWA fitting method and the extraction of production cross section a combined fit of
several experimental data sample provides a very powerful tool to extract the excitation function for
non-resonant and resonant production of pK+$\Lambda$ in p+p collisions, as well to determine the singlett
and triplett scattering length of p-$\Lambda$ S-Wave final state interaction.\\
In this paper the basic framework was explained, a fundamental test for the reliabilty of the extracted scattering length and the method to determine the total cross section were presented. The results are nescessary for the further work, which will be presented in near future.

\end{document}